\documentclass[reprint,aps,prb,groupedaddress,showpacs]{revtex4-1}
\usepackage{bm}
\usepackage{amssymb}
\usepackage{amsmath}
\usepackage{dcolumn}
\usepackage{threeparttable}
\usepackage{rotfloat}
\usepackage{float}
\usepackage{color}
\usepackage{booktabs}
\usepackage{epstopdf}

\begin{document}
\title{%
Filling-enforced Dirac loops and their evolutions under various perturbations
}

\author{Dexi Shao$^1$}
\email[]{sdx@iphy.ac.cn}
\author{Chen Fang$^{1}$}
\email[]{cfang@iphy.ac.cn}

\affiliation{
$^1$ Beijing National Laboratory for Condensed Matter Physics, and Institute of Physics, Chinese Academy of Sciences, Beijing 100190, China
}

\begin{abstract}
 Based on symmetry analysis, we propose that filling-enforced Dirac loops (FEDLs) in non-magnetic systems exist and only exist in only five space groups (SGs), namely, SG.57, SG.60, SG.61, SG.62 and SG.205. 
 We explore all possible configurations of the FEDLs in these space groups, and classify them accordingly. Furthermore, we study the evolutions of the FEDLs under various types of symmetry-breaking perturbations, such as an applied strain or an external field. The results show that FEDL materials can serve as parent materials of both topological semimetals hosting nodal points/loops, and topological insulators/topological crystalline insulators. By means of first-principles calculations, many materials possessing FEDLs are predicted.
\end{abstract}

\maketitle

\section{Introduction}
Topological materials have attracted great interest both theoretically and experimentally~\cite{TIs-Hasan2010, TIs-FuLiang2011, Armitage2018} since the proposal of topological insulators (TIs)~\cite{Kane2005}. Generally, topological materials can be classified into gapped systems and gapless systems according to the electronic states near the Fermi level. 
The most famous examples among them are TIs and Weyl semimetals (WSMs)~\cite{Wan2011,Xu2011,Yang2011, Burkov2011a, Burkov2011b, Halasz2012, Zyuzin2012, Lu2012, Das2013, Liu2014, zhang2014a, Weng2015, Xu2015a, Lv2015a,  Yang2015, Lv2015b, Xu2015b, Alidoust2015, huang2015, Xu2016, Lu2015, Ruan2016a, Ruan2016b}, respectively.
In terms of the gapped systems, the nontrivial topology of the bands can be characterized by a topological invariant which depends on the Bloch wave functions of all the occupied bands in the whole Brillouin zone (BZ).
And it is well-known that symmetries always play key roles in the classification of them. One of the celebrated examples is the ``periodic table'' of noninteracting TIs and topological superconductors (TSCs) characterized by time-reversal symmetry (TRS), particle-hole symmetry, and chiral symmetry~\cite{Schnyder2008, Kitaev2009}.
In addition, the crystal symmetries are found to give rise to a new kind of TIs, i.e., the topological crystalline insulators (TCIs)~\cite{class-TI-Tsc-2008, TCI-FuLiang-2011, SG-class-2013, TCI-SG-2013, TCI-TSC-2014, Wang2016Hourglass}. Recently, TCIs in non-magnetic systems have been enumerated~\cite{Songzd-TCI-2018,Khalaf-2018,Songzd-TCI-2019}, and these states can be fast-diagnosed by symmetry eigenvalues~\cite{Bradlyn2017,Ashvin-SI-2017,Vergniory2019,ZhangTT2019,TangF2019,Bernevig-band-discon-2018,Bernevig-band-graphtheory-2018,Bernevig-EBR-2018}.
Besides TIs and TCIs, many gapless topological phases have also been proposed, such as Dirac semimetals~\cite{Wang2012, Young2012, Wang2013,Liu2014A, Chen2014, Neupane2014, Nagaosa2014, Xu2015Observation, Xu2017, Huang2017, tang2016dirac, Wang2017Antiferromagetic,ZhangD2018, hua2018dirac,hex2018}, node-line semimetals~\cite{Kim2015Dirac,Yu2015Topological, Bian2016Drumhead, Fang2016Topological,  Yu2017Topological, Li-K3P4-2017, Sun2017Dirac,Chang2017},
nodal surface semimetals~\cite{BaMX3-Liang2016,2DGraphene-NS-Zhong2016,IS-NSSM-Bzuifmmode2017,2D-HfGeTe-Guan2017,NSSM-Wu-2018}, hopf-link semimetals~\cite{, Chen2017,Yan2017Nodal} and many other semimetals with unconventional quasiparticles beyond Dirac and Weyl fermions~\cite{Bradlyn2016Beyond,Wieder2016DDSM,Fang-SI-2018}. All these findings have greatly improved our knowledge of both the gapped and the gapless topological phases.

Guided by the compatibility relations~\cite{Bradlyn2017}, many nonsymmorphic-symmetry-enforced degeneracies have been proposed. Especially, there exists a new type of degeneracies which are filling-enforced. Fillings that realize a band insulator for all nonsymmorphic space groups (SGs) with and without SOC are listed~\cite{Watanabe-filling-2016}. Systems with other fillings must be gapless, namely, the filling-enforced semimetals. Until now, many filling-enforced semimetals have been proposed~\cite{Bradlyn2017,Vergniory2019,ZhangTT2019,TangF2019}, among which filling-enforced Dirac loop (FEDL) semimetal is a special example with four-fold degenerate nodal loops in the BZ. Different from WSMs which only require lattice translation symmetry for their protection,
materials with FEDL require some more symmetries. Although the FEDLs have been proposed in several systems~\cite{Kee-2012,Kee-2015,ReO2-2017,Li2018,Nam2019,Shao-AgF2-2019},
a general idea for searching the FEDL materials is still missing.

In this work, we first explore the necessary and sufficient conditions for FEDLs in non-magnetic systems, and find that there are five and only five SGs possessing the FEDLs. Then, we further explore all possible configurations of the Dirac loops and classify them accordingly. Motivated by earlier works~\cite{Levy2010,Shao2017}, we study the evolutions of the FEDLs under various perturbations, 
and find that the FEDL materials can serve as parent materials of both various topological semimetals and TIs/TCIs. At last, almost all the FEDL materials are listed, from which we have picked out some good candidates with fewer and smaller electron/hole pockets near the Fermi level by means of first-principles calculations. 

\section{Methods}
In the preparation phase, we import the non-magnetic half-filled (with the filling $8n+4$) materials in the corresponding SGs that are both registered in the online crystal database the Materials Project ($https:// materialsproject.org$)~\cite{MaterProject} and the Inorganic Crystal Structure Database (ICSD; $http://www2.fiz-karlsruhe.de/icsd \_home.html$)~\cite{ICSD}. By ``non-magnetic'', we regard one material as non-magnetic if its magnetic moment is not higher than 0.1 $\mu_B$ per unit cell (according to its Materials Project record).

We perform first-principles calculations based on the density functional theory (DFT) using projector augmented wave (PAW) method implemented in the Vienna ab initio simulation package (VASP)~\cite{kresse1999}. The generalized gradient approximation (GGA), as implemented in the Perdew-Burke-Ernzerhof (PBE) functional~\cite{GGA-PBE1996} is adopted to get the band structures. The cutoff parameter for the wave functions was 500 eV. The BZ was sampled by Monkhorst-Pack method~\cite{Monkhorst-BZ1976} with a k-spacing of $0.025\times 2\pi {\AA}^{-1}$ for the three-dimensional periodic boundary conditions.

\section{Results and Discussions}
\subsection{General descriptions of FEDLs}
\begin{figure}[bh]
\begin{center}
\includegraphics[width=0.48\textwidth]{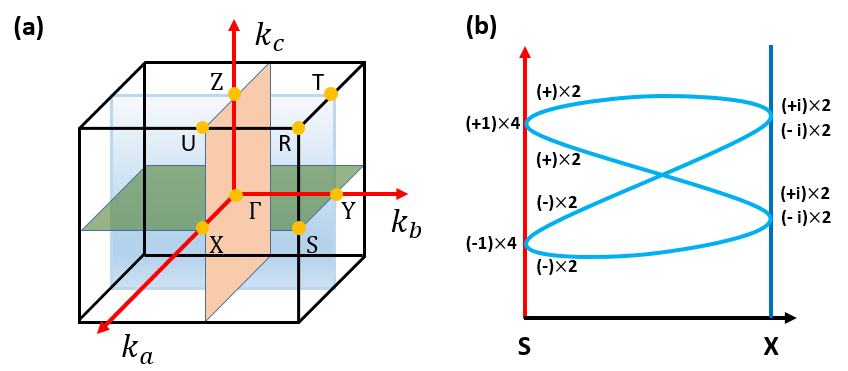}
\caption{%
  The schematic figures of (a) the corresponding BZ of SG.61. (b) the hourglass dispersion along S--X. Labels of the vertical axis in the right figure denote $\tilde{G}_{x}$ eigenvalue $g_{x}$. We can find the four-fold degenerate band crossings due to the hourglass dispersion protected by the $\tilde{G}_{x}$ symmetry between two quartets.
}
\label{fig:hourglass-sy}
\end{center}
\end{figure}

Up to now, there are some FEDL materials proposed in earlier works~\cite{ReO2-2017,Li2018,Nam2019,Shao-AgF2-2019}. Among them, $Pbca$ AgF$_2$ is a good example with the hourglass dispersion between S and X in the band structures, as shown in Fig.~\ref{fig:hourglass-sy}. It is the hourglass dispersion protected by the glide-mirror symmetry $\tilde{G}_{x}$ that contributes to the FEDL. Different from the hourglass dispersion in the nodal-chain metals~\cite{Bzdu2016Nodal}, the hourglass dispersion leading to the FEDL owns some unique features. The first feature is the existence of the TRS ($T$) and the inversion symmetry ($P$), which are required for the locally double degeneracy at each $\vec{k}$ point in the BZ. Secondly, the Dirac loop is always protected by some glide-mirror symmetry $\tilde{G}_{\alpha}$ and located on the $k_{\alpha}=\pi$ plane. Here, $\alpha= l x+m y+n z$ denotes the normal direction of the glide-mirror symmetry $\tilde{G}_{\alpha}$ with the Miller indices $\langle lmn\rangle$.

Here we give a general description of a Dirac loop. A Dirac loop is a type of four-fold degenerate nodal loop locating in a time-reversal invariant plane and is constructed by the band crossings of two $P*T$ related doublets with opposite mirror/glide-mirror eigenvalues. These band crossings along any path connecting two time-reversal invariant momenta (TRIMs) originate from the hourglass-like dispersion protected by the mirror/glide-mirror symmetry. We say one plane in the BZ is time-reversal invariant if the set of $\vec{k}$ points in this plane map to the same plane up to a reciprocal lattice vector under TRS.

In the following, we will deduce the necessary and sufficient conditions of a FEDL.
\begin{figure}[bh]
\begin{center}
\includegraphics[width=0.48\textwidth]{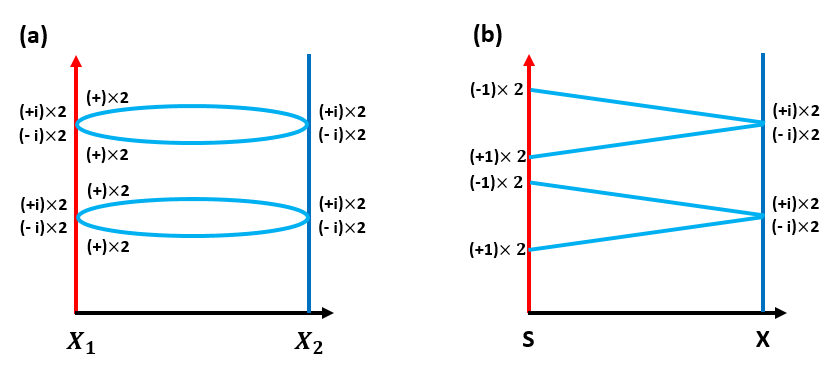}
\caption{%
  The schematic figures of band connections between (a) two `X'-type TRIMs and (b) one `S'-type TRIM and one `X'-type TRIM.
}
\label{fig:com-nec}
\end{center}
\end{figure}
\begin{itemize}
\item The combination of $P$ and $T$, i.e., $P*T$, gives the local double degeneracy at each $\vec{k}$ point in the BZ, termed as $P*T$ related doublets.
\item The existence of a mirror/glide-mirror symmetry $\tilde{G}_{\alpha}=\{m_{\alpha}|\vec{\tau}_{\alpha}\}$ serves each $P*T$ related doublet with the same $\tilde{G}_{\alpha}$ eigenvalues. Following the work by Fang \emph{et al.}~\cite{Fang2015}, $P*T$ related states sharing the same $\tilde{G}_{\alpha}$ eigenvalues appear only if the component of $\vec{\tau}_{\alpha}$ along the $\alpha$ direction is non-zero. Simultaneously, the Dirac loop can only exist in the $k_{\alpha}=\pi$ plane, while not the $k_{\alpha}=0$ plane. It can be seen from the following.
    In the $\tilde{G}_{\alpha}$ invariant plane, we have
\begin{equation}\label{eq:g}
\begin{split}
g_{\alpha}=\pm ie^{-i\vec{k}_{n\alpha}\cdot\vec{\tau}_{n\alpha}}.\\
\end{split}
\end{equation}
Here, $\vec{k}_{n\alpha}$ and $\vec{\tau}_{n\alpha}$ denote the component along directions other than $\alpha$ of $\vec{k}$ and $\vec{\tau}_{\alpha}$, respectively.
Suppose $|\psi\rangle$ is the eigenstate of $\tilde{G}_{\alpha}$ with $g_{\alpha}=\pm ie^{-i\vec{k}_{n\alpha}\cdot\vec{\tau}_{n\alpha}}$, then
\begin{equation}\label{eq:geig}
\begin{split}
\tilde{G}_{\alpha}(P*T)|\psi\rangle&=e^{-i(\vec{k}\cdot\vec{\tau}_{\alpha})}P\tilde{G}_{\alpha}T|\psi\rangle\\
                        &=e^{-i2(k_{\alpha}\tau_{\alpha}+\vec{k}_{n\alpha}\cdot\vec{\tau}_{n\alpha})}P\tilde{G}_{\alpha}T|\psi\rangle\\
                        &=e^{-i2(k_{\alpha}\tau_{\alpha}+\vec{k}_{n\alpha}\cdot\vec{\tau}_{n\alpha})}PT\tilde{G}_{\alpha}|\psi\rangle\\
                        &=e^{-i2(k_{\alpha}\tau_{\alpha}+\vec{k}_{n\alpha}\cdot\vec{\tau}_{n\alpha})}PTg_{\alpha}|\psi\rangle\\
                        &=g_{\alpha}^{*}e^{-i2(k_{\alpha}\tau_{\alpha}+\vec{k}_{n\alpha}\cdot\vec{\tau}_{n\alpha})}PT|\psi\rangle\\
                        &=\mp ie^{-i(2k_{\alpha}\tau_{\alpha}+\vec{k}_{n\alpha}\cdot\vec{\tau}_{n\alpha})}PT|\psi\rangle,\\
\end{split}
\end{equation}
which indicates $P*T|\psi\rangle$ is the eigenstate of $\tilde{G}_{\alpha}$ with $g_{\alpha}=\mp ie^{-i(2k_{\alpha}\tau_{\alpha}+\vec{k}_{n\alpha}\cdot\vec{\tau}_{n\alpha})}$. The requirement $P*T$ related doublets sharing the same $\tilde{G}_{\alpha}$ eigenvalue means
\begin{equation}\label{eq:geig}
\begin{split}
 ie^{-i\vec{k}_{n\alpha}\cdot\vec{\tau}_{n\alpha}}=- ie^{-i(2k_{\alpha}\tau_{\alpha}+\vec{k}_{n\alpha}\cdot\vec{\tau}_{n\alpha})},\\
\end{split}
\end{equation}
which requires $e^{-i(2k_{\alpha}\tau_{\alpha})}=-1$. This can be obtained only if $\tau_{\alpha}$ is a half-integer translation and $k_{\alpha}=\pi$.
\item There must exist another direction along which the component of $\vec{\tau}_{\alpha}$ is non-zero. This can be deduced from the following.
    If all the other components of $\vec{\tau}_{\alpha}$ along the other directions are zero, we have $\{\tilde{G}_{\alpha}, P\}=0$ and $g_{\alpha}=2\times (+i)\oplus 2\times (-i)$ for all the four TRIMs in the $k_{\alpha}=\pi$ plane. We call this type of TRIMs as `X'-type TRIMs. Band structures between two `X'-type TRIMs can be gapped, and the hourglass dispersion is no longer necessary to appear, such as the band connections shown in Fig.~\ref{fig:com-nec}(a). As a result, there exists another direction along which the component of $\vec{\tau}_{\alpha}$ is non-zero, which gives another type of TRIMs with $g_{\alpha}=2\times (\pm 1)$, termed as `S' type TRIMs. Thus, $\tilde{G}_{\alpha}$ here must be a glide-mirror symmetry, while can not be a mirror symmetry.
\item The three points above indicate that there exist two `X'-type TRIMs with $g_{\alpha}=2\times (+i)\oplus 2\times (-i)$ and two `S'-type TRIMs with $\tilde{g}_{\alpha}=2\times (\pm 1)$. Given no other symmetries, the hourglass dispersion between `S'-type TRIM and `X'-type TRIM is not necessary to appear, such as the counter example shown in Fig.~\ref{fig:com-nec}(b). Thus, another symmetry $\hat{R}$ is needed to introduce four-fold degeneracy with $\tilde{g}_{\alpha}=4\times (\pm 1)$ at one `S'-type TRIM. $g_{\alpha}=4\times (\pm 1)$ indicates that there exist two types of irreducible representations in which we have $D(\tilde{G}_{\alpha})= I_{4\times 4}$ or $D(\tilde{G}_{\alpha})= -I_{4\times 4}$ if some special quaternate bases are chosen. $D(\tilde{G}_{\alpha})$ and $I_{4\times 4}$ denote the representation matrix of $\tilde{G}_{\alpha}$ and the $4\times 4$ identity matrix. It is clearly that $[D(\tilde{G}_{\alpha}),D(\hat{R})]\equiv 0$ stands in every irreducible representation, regardless of which type of irreducible representation it belongs to. Thus, the introduced symmetry $\hat{R}$ should satisfy the following conditions
\begin{equation}\label{eq:con-fedl}
[\tilde{G}_{\alpha},\hat{R}]=0~,~\{\hat{R},P\}=0
\end{equation}
at this `S'-type TRIM. $\{\hat{R},P\}=0$ is required by the following reasons. Firstly, it is known that either $\{\hat{R},P\}=0$ or $[\hat{R},P]=0$ stands at any TRIM. Since $P\ast\hat{R}$ and $\hat{R}\ast P$ always share the same rotation component in the real space, the difference between them must be an integer translation $T_{r}$. Thus, $\hat{R}P=T_{r}P\hat{R}=e^{-i\vec{k}\cdot\vec{\tau}_{r}}P\hat{R}=\pm P\hat{R}$ stands when $\vec{k}$ is a $\hat{R}$-invariant TRIM for systems with symmetries $\hat{R}$ and $P$. $\vec{\tau}_{r}$ here denotes the translational vector of $T_{r}$. As a result, $\{\hat{R},P\}=0$ or $[\hat{R},P]=0$ stands at arbitrary TRIMs.
Furthermore, four-fold degeneracy can be induced only if $\{\hat{R},P\}=0$, while not $[\hat{R},P]=0$. In addition, $[\tilde{G}_{\alpha},\hat{R}]=0$ will constrain the four-fold degeneracy with $D(\tilde{G}_{\alpha})= I_{4\times 4}$ or $D(\tilde{G}_{\alpha})= -I_{4\times 4}$ in the quaternate bases $\{|\psi\rangle,P*T|\psi\rangle,\hat{R}|\psi\rangle,\hat{R}P*T|\psi\rangle\}$.

\end{itemize}
Given all these conditions above, hourglass Dirac dispersion protected by the $\tilde{G}_{\alpha}$ between `S'-type TRIM and `X'-type TRIM is constructed. Furthermore, nodal phenomena protected by some glide-mirror symmetry must be a one-dimensional nodal loop. Thus, the Dirac loop is induced.

There exist five and only five SGs satisfying these conditions above, namely, SG.57, SG.60, SG.61, SG.62 and SG.205.
Furthermore, we find that the fillings which realize a band insulator for all these groups are $8n$, and the Dirac loop comes from the band crossings between $P*T$ related doublets $8n+3\oplus 8n+4$ and $P*T$ related doublets $8n+5\oplus 8n+6$, suggesting these Dirac loops appear at the Fermi level when systems of these SGs are half-filled. As a result, we call these systems FEDL materials.

\subsection{Configurations of FEDLs in SG.57,60,61,62,205}
In the following, we focus on the configurations of the FEDLs in all these SGs, among which SG.60~\cite{ReO2-2017}, SG.61~\cite{Shao-AgF2-2019} and SG.62~\cite{Li2018,Nam2019} have been proposed in the earlier works.

\begin{figure}[bh]
\begin{center}
\includegraphics[width=0.48\textwidth]{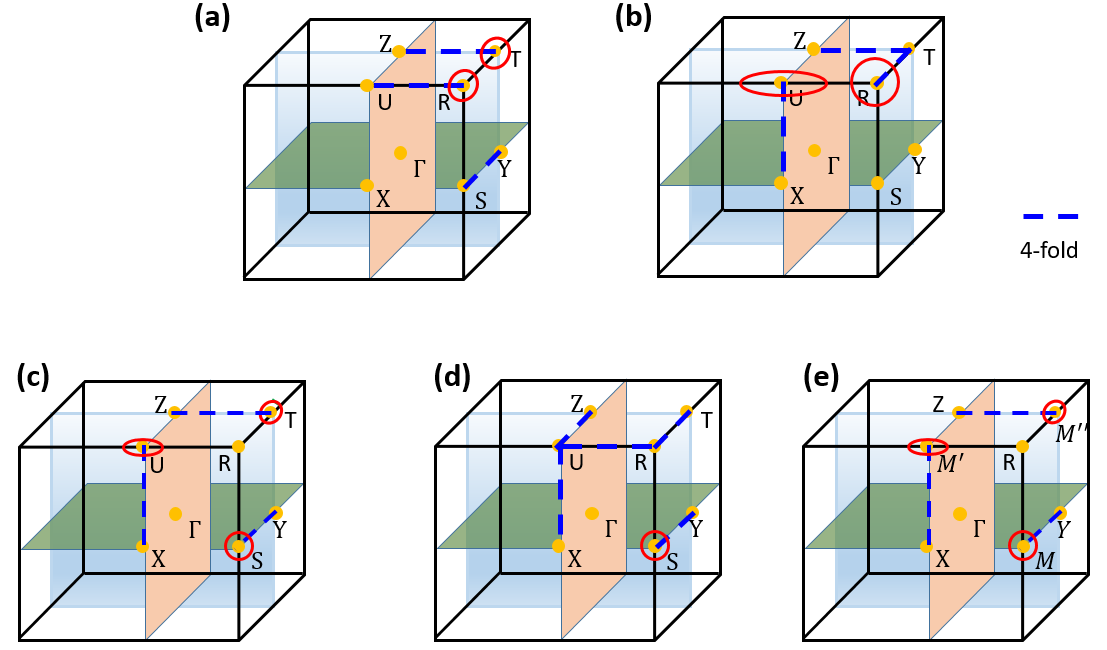}
\caption{%
  The schematic figures for the configurations of FEDLs in (a) SG.57, (b) SG.60, (c) SG.61, (d) SG.62 and (e) SG.205 in the BZ, respectively. The dashed blue lines here and in the following denote four-fold degenerate bands from the corresponding compatibility relations along these lines.
}
\label{fig:sg57-60-61-62-205}
\end{center}
\end{figure}

 According to the above discussion, we deduce the schematic figures of all the FEDLs for the five SGs, as shown in Fig.~\ref{fig:sg57-60-61-62-205}. We have chosen the cases that all the Dirac loops are separated and confined in the BZ for simplicity in the figure. The words ``confined in the BZ'' (``traverse the BZ'') mean that the Dirac loop can (can not) be translated to the first BZ without touching any boundary of the BZ. However, these Dirac loops can touch with each other and traverse the BZ~\cite{ReO2-2017,Shao-AgF2-2019}. It is natural for us to explore all the possible configurations of the FEDLs and try to classify them.

\subsubsection{Configurations of FEDLs in SG.57}
\begin{figure}[bh]
\begin{center}
\includegraphics[width=0.40\textwidth]{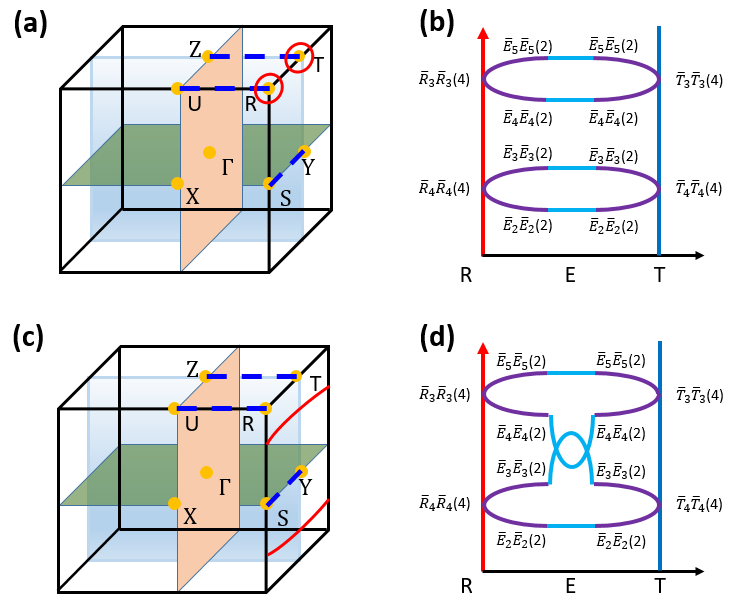}
\caption{%
  The schematic figures for the configurations of FEDLs in SG.57 in the BZ. The two separated red rings in figure (a) and two red lines traversing the BZ in figure (c) denote the separated FEDLs and FEDLs traversing the BZ, respectively. The right panels show the corresponding connections of bands satisfying the compatibility relations along R-T.
}
\label{fig:sg57-con}
\end{center}
\end{figure}

As shown in Fig.\ref{fig:sg57-con}(a) and Fig.\ref{fig:sg57-con}(c), there are two different configurations of FEDLs in SG.57, named as separated FEDLs and FEDLs traversing the BZ. The corresponding connections of the bands satisfying the compatibility relations are shown in Fig.\ref{fig:sg57-con}(b) and Fig.\ref{fig:sg57-con}(d), respectively.
According to the compatibility relation along R-T, we emphasize here that Dirac chain composed of the two FEDLs touching at some isolated point along R-T is prohibited. 

\subsubsection{Configurations of FEDLs in SG.60}
There are two FEDLs in SG.60, with one lying in $k_{x}=\pi$ plane, while the other lying in $k_{z}=\pi$ plane. They are protected by $\tilde{G}_{x}$ and $\tilde{G}_{z}$, respectively.

\begin{figure*}[th]
\begin{center}
\includegraphics[width=0.75\textwidth]{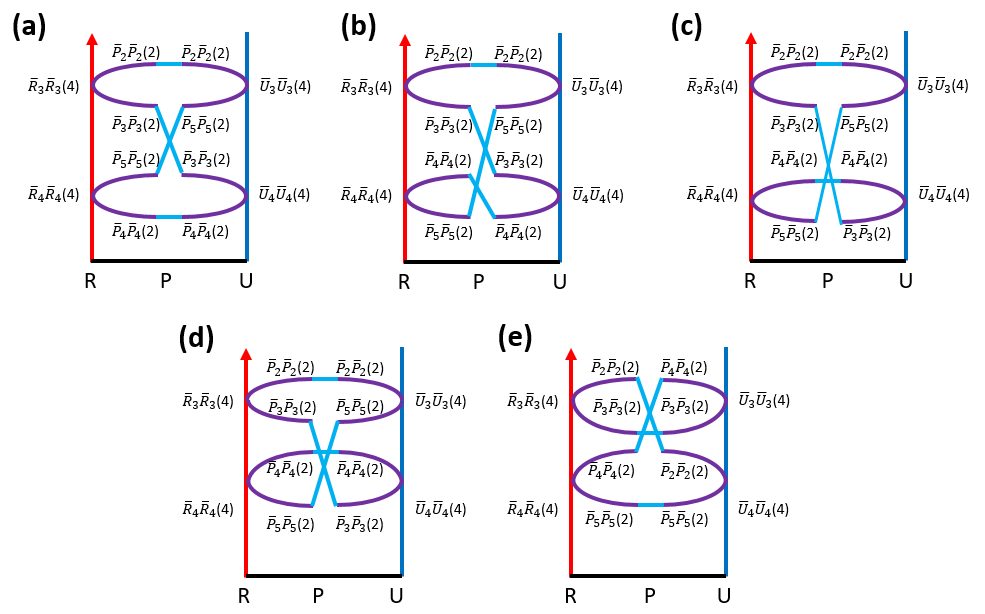}
\caption{%
  (a-e) Five possible connections of the bands satisfying the compatibility relations along R-U in SG.60.
}
\label{fig:sg60-con}
\end{center}
\end{figure*}

To obtain some intuitive pictures, we list several possible band connections according with the compatibility relation along U-R, which gives a Dirac chain from Figs.~\ref{fig:sg60-con}(a-c), and two separated FEDLs from Figs.~\ref{fig:sg60-con}(d-e), respectively. Reminding of the earlier work~\cite{Chang2017}, one may consider a third configuration of the FEDLs, i.e., the Hopf Dirac links.
 The hypothetical Hopf Dirac link appears when these two separated FEDLs go close, and then cross with each other. Thus, the crossing point of the Hopf Dirac link near R is protected by $\tilde{G}_{z}$, while the crossing point near U is protected by $\tilde{G}_{x}$. Band crossings protected by $\tilde{G}_{z}$ can only come from bands between $\bar{P}_{2}\bar{P}_{2}(2)(\bar{P}_{4}\bar{P}_{4}(2))$ and $\bar{P}_{3}\bar{P}_{3}(2)(\bar{P}_{5}\bar{P}_{5}(2))$. However, both bands $\bar{P}_{2}\bar{P}_{2}(2)(\bar{P}_{4}\bar{P}_{4}(2))$ and $\bar{P}_{3}\bar{P}_{3}(2)(\bar{P}_{5}\bar{P}_{5}(2))$ are simultaneously valence bands or conductance bands, as a result, Hopf Dirac link can be obtained by bands below or above the Fermi level, while can not be obtained by bands between $8n+3\oplus8n+4$ and $8n+5\oplus8n+6$ at the Fermi level.

In addition, the Dirac loop encircling R in the $k_{x}=\pi$ plane is either traversing or confined in the BZ along $k_z$ axis. We define a new $\zeta_{2}$ index as
\begin{equation}\label{eq:con-fedl}
\zeta_{2}= \frac{n_{+}(S)-n_{+}(R)}{2}~mod~2
\end{equation}
to tell whether this Dirac loop is traversing the BZ or not. In Eq.~(\ref{eq:con-fedl}), $n_{+}(R)$ and $n_{+}(S)$ represent the number of occupied bands with $+1$ $\tilde{G}_{x}$ eigenvalue at R and S, respectively. $\zeta_{2}=0$ indicates that this Dirac loop goes traversing the BZ, while $\zeta_{2}=1$ indicates this Dirac loop is confined in the BZ.

\subsubsection{Configurations of FEDLs in SG.61}
\begin{figure}[bh]
\begin{center}
\includegraphics[width=0.45\textwidth]{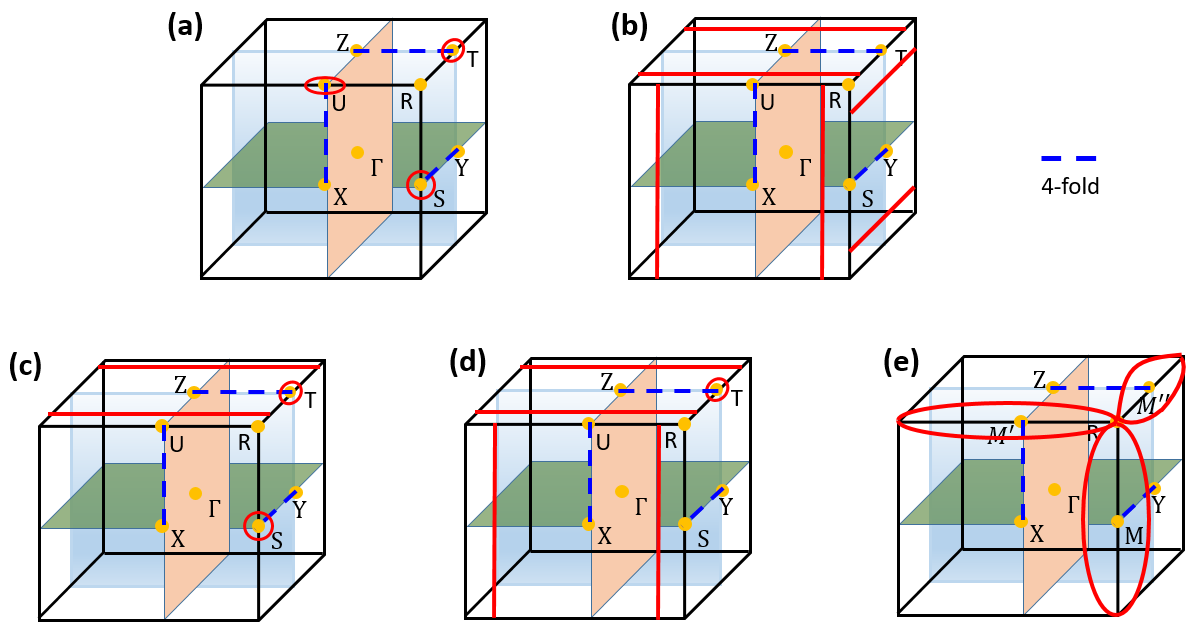}
\caption{%
  (a,b,c,d) The schematic figures for the configurations of FEDLs in SG.61 in the BZ. We have neglected the possible connections of different FEDLs for simplicity.
  (a,b,e) The schematic figures for the configurations of FEDLs in SG.205 in the BZ.
}
\label{fig:sg61-205}
\end{center}
\end{figure}

SG.61 (\emph{Pbca}) contains three screw axes, the inversion and three glide mirrors, as shown in Table.~\ref{table:space-group}. The subscripts satisfy $(\alpha,\beta)=\{(x,y);(y,z);(z,x)\}$, which presents a cyclic permutation relation. As a result, there exists a FEDL on each surface of the BZ surrounding S, T and U, respectively. Furthermore, each FEDL can be either traversing or confined in the BZ, which can be distinguished by three $(\zeta_{2x},\zeta_{2y},\zeta_{2z})$ indexes
\begin{equation}\label{eq:SG61-zxyz}
\begin{split}
&\zeta_{2x}=\frac{n^{x}_{+}(S)-n^{x}_{+}(R)}{2}~mod~2,\\
&\zeta_{2y}=\frac{n^{y}_{+}(T)-n^{y}_{+}(R)}{2}~mod~2,\\
&\zeta_{2z}=\frac{n^{z}_{+}(U)-n^{z}_{+}(R)}{2}~mod~2.\\
\end{split}
\end{equation}
In Eq.~(\ref{eq:SG61-zxyz}), $n^{x}_{+}(S)/n^{y}_{+}(T)/n^{z}_{+}(U)$ and $n^{x/y/z}_{+}(R)$ represent the numbers of occupied bands with $+1$ $\tilde{G}_{x/y/z}$ eigenvalue at $S/T/U$ and R, respectively. $\zeta_{2\alpha}=0$ indicates the Dirac loop in the $k_{\alpha}=\pi$ plane traverses the BZ, while $\zeta_{2\alpha}=1$ indicates this Dirac loop is confined in the BZ. As a result, FEDLs shown in Figs.~\ref{fig:sg61-205}(a), (b), (c) and (d) indicate the corresponding $\{\zeta_{2x},\zeta_{2y},\zeta_{2z}\}$ equals $\{1,1,1\}$, $\{0,0,0\}$, $\{1,1,0\}$ and $\{1,0,0\}$, respectively.

It should be noted that when $\{\zeta_{2\alpha},\zeta_{2\beta}\}=\{0,1\}$ for some system, the Dirac loops in the $k_{\alpha}=\pi$ and $k_{\beta}=\pi$ planes may touch each other, and thus, a Dirac chain forms. Such as the $Pbca$ AgF$_2$ system~\cite{Shao-AgF2-2019}. 

\begin{table}
\caption{The operators in SG \emph{Pbca}.}
   \begin{tabular}{l r}%
       \hline\hline
          \textbf{Operators}        & \textbf{Symmetry representation}         \\
            $E$                       & \{$1\mid 0$\}          \\
           $\tilde{R}_{2\alpha}$       & \{$R_{2\alpha}|\frac{1}{2}(a_{\alpha}+a_{\beta})$\} \\
            $P$                       & \{$-1\mid 0$\} \\
           $\tilde{M}_{\alpha}$       & \{$M_{\alpha}|\frac{1}{2}(a_{\alpha}+a_{\beta})$\} \\
       \hline\hline
  \end{tabular}
\label{table:space-group}
\end{table}

\subsubsection{Configurations of FEDLs in SG.62}
There is only one FEDL in the $k_{x}=\pi$ plane, and it must be confined in the BZ because there is only one four-dimensional irreducible representation along X-U-R of SG.62. Thus, bands between $8n+3\oplus8n+4$ and $8n+5\oplus8n+6$ must be gapped along X-U-R, leading the FEDL confined in the BZ.

\subsubsection{Configurations of FEDLs in SG.205}
SG.205 can be seen as a cubic case of SG.61 as a result of SG.205=SG.61$\otimes R_{3[111]}$. $R_{3[111]}$ denotes the three-fold rotation along the body-diagonal direction. Thus, these three FEDLs are simultaneously either traversing the BZ or confined in the BZ, and related with each other by the $R_{3[111]}$ symmetry, as shown in Fig.~\ref{fig:sg61-205}(a) and Fig.~\ref{fig:sg61-205}(b), respectively. More interestingly, we find that the separated FEDLs may connect with each other at R, which forms a Dirac nodal net, as schematized in Fig.~\ref{fig:sg61-205}(e). \par
Similar with SG.60 and SG.61, we can introduce $\zeta_{2}=\zeta_{2x}=\zeta_{2y}=\zeta_{2z}$ as
\begin{equation}\label{eq:SG205}
\begin{split}
&\zeta_{2}=\frac{n^{x}_{+}(M)-n^{x}_{+}(R)}{2}~mod~2\\
\end{split}
\end{equation}
to distinguish the three different configurations in SG.205. $\zeta_{2}=1$ indicates that the three FEDLs are confined in the BZ, while $\zeta_{2}=0$ indicates the three FEDLs traverse the BZ or just connect with each other at R. To further identify whether the FEDLs are composed of three separated Dirac loops traversing the BZ or connected with each other at R when $\zeta_{2}=0$, more calculations along M-R are needed. In the following, we will use IrN$_2$ as an example to illustrate this.

\begin{figure*}[th]
\begin{center}
\includegraphics[width=1.0\textwidth]{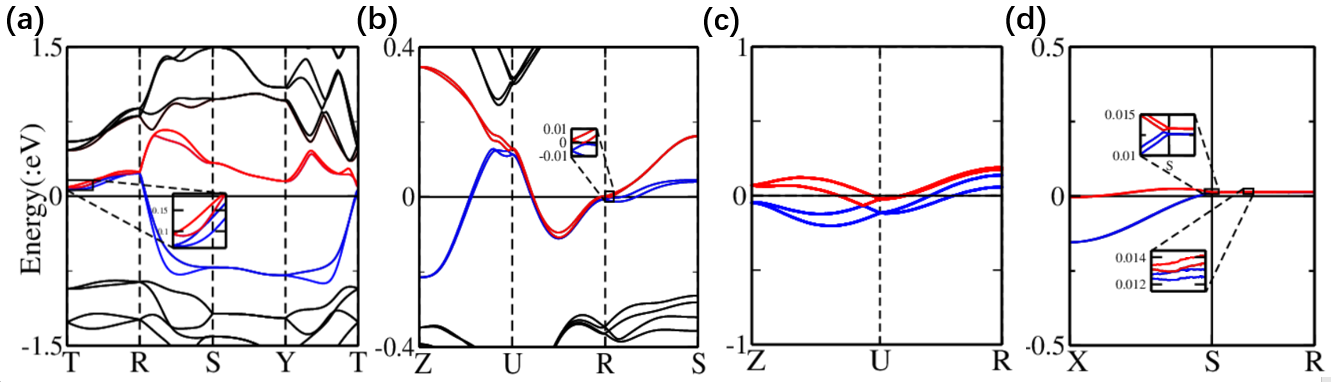}
\caption{%
  Band structures of (a) $Pbcm$ K$_2$SnBi, (b) $Pbcn$ Hf$_2$Co3Si4, (c) $Pbca$ AgF$_2$ and (d) $Pnma$ AgAsF$_7$, respectively. The insets in them show us the hourglass dispersions protected by the corresponding glide-mirror symmetries in SG.57, SG.60, SG.61 and SG.62, respectively.
}
\label{fig:sg57-62-band}
\end{center}
\end{figure*}

\begin{figure}[bh]
\begin{center}
\includegraphics[width=0.5\textwidth]{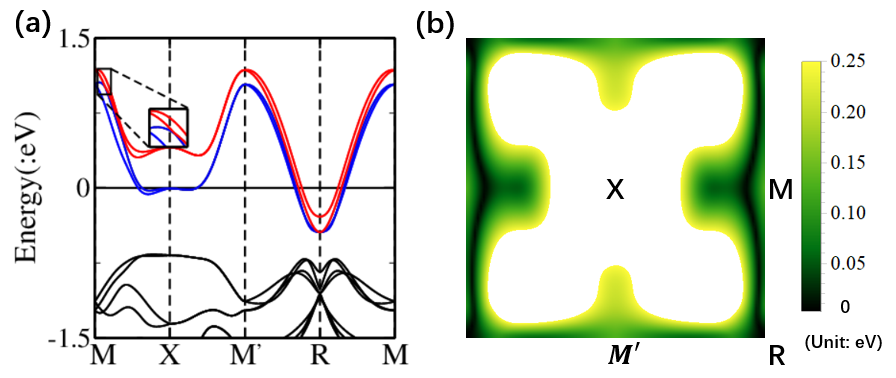}
\caption{%
  (a) Band structures of $Pa\bar{3}$ IrN$_2$.
  (b) Illustration figure of the direct gap between the minimum conduction band and the maximum valence band near the Fermi level on the $k_{x}=\pi$ plane. We choose 0.25 eV as the energy cut, i.e., zone with energy gap larger than 0.25 eV is represented with the white color.
}
\label{fig:sg205-band}
\end{center}
\end{figure}

To justify the FEDLs in each SG, we calculate band structures of $Pbcm$ K$_2$SnBi, $Pbcn$ Hf$_2$Co$_3$Si$_4$, $Pbca$ AgF$_2$, $Pnma$ AgAsF$_7$ and $Pa\bar{3}$ IrN$_2$, as shown in Figs.~\ref{fig:sg57-62-band}(a-d) and Figs.~\ref{fig:sg205-band}(a) and (b), respectively. We can see the hourglass dispersions protected by the corresponding glide-mirror symmetry from the band structures in each SG, which indicates they are indeed FEDL materials. Especially, for the case of $Pa\bar{3}$ IrN$_2$, the FEDL lies close to M-R and touches this line at R, as the black line shows in Fig.~\ref{fig:sg205-band}(b). It means that the corresponding FEDL is no other than the Dirac nodal net, as schematized in Fig.~\ref{fig:sg61-205}(e).

\section{Evolutions of FEDLs under symmetry-breaking but translation-invariant transitions}
In the above, we have deduced all possible configurations of the FEDLs in the five SGs, and classify them accordingly. Furthermore, it is well-known that perturbations like strains, external magnetic fields, circularly polarized lights and so on, can serve as useful methods to tune the band structures in condensed matter systems. Generally, these perturbations will break some symmetries, and then, how the FEDLs in the five SGs evolve under these perturbations remains a question. In the following, we will explore the evolutions of these FEDLs under maximal-subgroup symmetry-breaking transitions which respect the translation symmetries. Suppose all the transitions are obtained by the adiabatic perturbations, which indicates that no new band crossings occur in this progress. Furthermore, it should be noted that we have neglected what and how the perturbations are applied to get the corresponding transitions.

Firstly, taking SG.57 as an example, we can get all types of symmetry-breaking transitions of SG.57 from the Bilbao Crystallographic Server. We sign the transition as SG.57-sg.n, where ``sg.n'' denotes some maximal subgroup of SG.57 with k-index = 1 (k-index = 1 indicates that the corresponding transition does not break any translation symmetry).
\begin{table*}
\caption{The symmetry-breaking transitions of SG.57.}
   \begin{tabular}{c c c c}%
       \hline\hline
          \textbf{SG.57-sg.n} & \textbf{Filling factor of sg.n}    & \textbf{Coordinate transformation}      & \textbf{Phase transition}   \\
            sg.11             &  4n                                 & $\{a,b,c\}_{sg.11}=\{b,c,a\}_{SG.57}$     & \{SFEDLs~,~FEDLTBZs\}$\rightarrow$ I\\
            sg.13             &  4n    & $\{a,b,c\}_{sg.13}=\{c,a,b\}_{SG.57}$     & \{SFEDLs~,~FEDLTBZ\}$\rightarrow$ \{$2\mathcal{DP}\oplus 2\mathcal{DP}$~,~I\}\\
            sg.14             &  4n                                 & $--$                                      & \{SFEDLs~,~FEDLTBZs\}$\rightarrow$ \{SDNLs~,~DNLTBZs\}\\
            sg.18             &  4n    & $\{a,b,c\}_{sg.18}=\{b,c,a\}_{SG.57}$     & \{SFEDLs~,~FEDLTBZs\}$\rightarrow$ \{ $2\mathcal{DN}\oplus 2\mathcal{DN}$~,~I\}\\
            sg.26             &  4n                                 & $\{a,b,c\}_{sg.26}=\{c,a,b\}_{SG.57}$     &   FEDLs$\rightarrow$ I\\
            sg.28             &  4n                                 & $\{a,b,c\}_{sg.28}=\{c,b,-a\}_{SG.57}$    & \{SFEDLs~,~FEDLTBZs\}$\rightarrow$\{4 CWNLs~,~4 WNLTBZs\}\\
            sg.29             &  8n                                 & $\{a,b,c\}_{sg.29}=\{-b,a,c\}_{SG.57}$    & \{SFEDLs~,~FEDLTBZs\}$\rightarrow$\{2 CWNCs~,~4 WNLTBZs\}\\
       \hline\hline
  \end{tabular}
  \begin{tablenotes}
        \footnotesize
        \item[1] The ``Filling factor of sg.n'' column denotes fillings that realize a band insulator for the subgroup sg.n.
        \item[2] SFEDLs and FEDLTBZs denote the separated FEDLs and the FEDLs traversing the BZ, respectively. Similarly, SDNLs and DNLTBZs represent separated Dirac nodal loops and Dirac nodal loops traversing the BZ, respectively.
        \item[3] $2\mathcal{DP}\oplus 2\mathcal{DP}$ and $2\mathcal{DN}\oplus 2\mathcal{DN}$ mean two Dirac points and two DNs along the R-T line are possible to exist when each separated FEDL is protected by $\tilde{R}_{2x}$ symmetry. DN here represents a four-fold degenerate node along R--T, with its dispersions doubly degenerate in the $k_{y}=\pi$ and $k_{z}=\pi$ planes, while without degeneracy along the other directions.
        \item[4]I, CWNL, WNLTBZ and CWNC denote insulator, concentric Weyl nodal loop, Weyl nodal loops traversing the BZ and concentric Weyl nodal chain, respectively.
        \item[5]Here ``$A \oplus B$'' indicates nodal phenomena $A$ and $B$ are independent with each other, and can be obtained simultaneously.
      \end{tablenotes}
\label{table:pt-SG57}
\end{table*}
If the FEDL crosses the R-T line, band crossings from the FEDL along this line must be simultaneously protected by either the $\tilde{R}_{2x}$ or the $\tilde{G}_{z}$ symmetry (besides the $\tilde{G}_{y}$ symmetry) from the compatibility relation.
Furthermore, if the separated FEDLs in the $k_{y}=\pi$ plane is protected by the $\tilde{G}_{z}$ along R-T, the separated FEDLs are in fact separated Dirac chains.
It should be noted that for the case of separated Dirac nodal chain, it can be divided into two parts, i.e., the FEDL in $k_{y}=\pi$ plane and the accidental Dirac nodal loop in $k_{z}=\pi$ plane. The evolutions of them can be deduced independently. In the following discussions, we will only focus on the evolution of FEDLs,
while the evolution of the accidental Dirac nodal loop in the $k_{z}=\pi$ plane is neglected.

\subsection{SG.57-sg.11 transition}
The corresponding perturbation of SG.57-sg.11 transition breaks $\tilde{R}_{2x/y}$ and $\tilde{G}_{x/y}$ symmetries, while $\tilde{R}_{2z}$, $\tilde{G}_{z}$ and the inversion symmetries are reserved.
Fillings that realize a band insulator for subgroup sg.11 is $4n$, which indicates the symmetry-enforced nodal phenomena of the subgroup are within bands $4n$---$4n+3$.
These symmetry-enforced nodal phenomena are different from the FEDLs within bands $8n+3$---$8n+6$ at the Fermi level.
As a result, whether these systems become insulators or topological semimetals depends on how the FEDLs evolve. Furthermore, the key for the evolution of the FEDLs is to tell whether the FEDLs are protected by the preserved symmetries at the corresponding high-symmetry $\vec{k}$ points. The reason is direct, the degeneracy from the FEDL will be gapped if it is no longer protected by any reserved symmetry.
\begin{itemize}
\item $\tilde{R}_{2z}$ invariant lines, R--S and T--Y: From the corresponding compatibility relations, $P*T$ related states possess the opposite $\tilde{R}_{2z}$ value. Then, there exist only one irreducible representation along these two lines in sg.11, which indicates FEDLs along these lines will be gapped under SG.57-sg.11 transition.
\item $\tilde{G}_{z}$ invariant line, R--T: When the FEDLs are separated, they must go across R--T. From the corresponding compatibility relation, we known $P*T$ related states along this line share the same $\tilde{G}_{z}$ value. Then, there exist two irreducible representations signed with different $\tilde{G}_{z}$ values along R--T in sg.11. Thus, band crossings from the FEDLs along R--T can be protected by $\tilde{G}_{z}$, which indicates that the FEDLs may be separated Dirac chains. The part of accidental Dirac loops (if exist) protected by $\tilde{G}_{z}$ in the $k_{z}=\pi$ plane is reserved, while the part of FEDL in the $k_{y}=\pi$ plane is gapped under SG.57-sg.11 transition.
\end{itemize}
In summary, the FEDLs in $k_{y}=\pi$ plane will be gapped, while the accidental Dirac nodal loops in $k_{z}=\pi$ plane (if exist) are reserved under the SG.57-sg.11 transition. The corresponding phase transitions induced by symmetry-breaking perturbations are listed in Table~\ref{table:pt-SG57}.

\subsection{SG.57-sg.14 transition}
The corresponding perturbation breaks $\tilde{R}_{2x}$, $\tilde{R}_{2z}$, $\tilde{G}_{x}$ and $\tilde{G}_{z}$ symmetries, while the $\tilde{R}_{2y}$, $\tilde{G}_{y}$ and inversion symmetries are reserved. In this case, the FEDLs remain to be accidental Dirac nodal loops because the reserved symmetries are sufficient for Dirac nodal loops in the $k_{y}=\pi$ plane. However, it should be noted that the Dirac nodal loops are no longer filling-enforced. Thus, the separated FEDLs and FEDLs traversing the BZ will evolve into separated Dirac nodal loops and Dirac nodal loops traversing the BZ, respectively.

\subsection{SG.57-sg.28 transition}
The corresponding perturbation breaks the $\tilde{R}_{2y}$, $\tilde{R}_{2z}$, $\tilde{G}_{x}$ and inversion symmetries, while $\tilde{R}_{2x}$, $\tilde{G}_{y}$ and $\tilde{G}_{z}$ symmetries are reserved. The absence of the inversion symmetry will break the double degeneracy induced by $P*T$, thus, each FEDL will split into two Weyl nodal loops. As a result, for the case of separated FEDLs, each FEDL evolves into two concentric Weyl nodal loops, while for the case of FEDL traversing the BZ, each FEDL evolves into two Weyl nodal loops traversing the BZ.

\subsection{SG.57-sg.29 transition}
Similar with the SG.57-sg.28 transition, each FEDL will split into two Weyl nodal loops as a result of the breaking of the inversion symmetry. Furthermore, there exists $\Theta=\tilde{G}_{y}*T$ enforced double degeneracy along the R--T line, which indicates that for the case of separated FEDLs, the concentric Weyl nodal loops from the same FEDL touch each other along the R-T line, leading to two concentric Weyl nodal chains.

\begin{table*}
\caption{The symmetry-breaking transitions of SG.60.}
   \begin{tabular}{c c c c}%
       \hline\hline
          \textbf{SG.60-sg.n} & \textbf{Filling factor of sg.n}    & \textbf{Coordinate transformation}      & \textbf{Phase transition}   \\
            sg.13             &  4n                                 & $--$                                      & \{SFEDL~,~FEDC\}$\rightarrow$ \{4 DPs~,~I\}\\
            sg.14-class a     &  4n                                & $\{a,b,c\}_{sg.14}=\{c,a,b\}_{SG.60}$     & \{SFEDL~,~FEDC\}$\rightarrow$  DNLx\\
            sg.14-class b     &  4n                                 & $\{a,b,c\}_{sg.14}=\{a,c,-a-b\}_{SG.60}$  & \{SFEDL~,~FEDC\}$\rightarrow$  DNLz\\
            sg.18             &  4n                                 & $\{a,b,c\}_{sg.18}=\{c,a,b\}_{SG.60}$     & \{SFEDL~,~FEDC\}$\rightarrow$ \{4 DNs~,~I\}\\
            sg.29             &  8n                                 & $\{a,b,c\}_{sg.29}=\{-b,a,c\}_{SG.60}$    & \{SFEDL~,~FEDC\}$\rightarrow$  CWNCx\\
            sg.30             &  4n                                 & $\{a,b,c\}_{sg.30}=\{c,a,b\}_{SG.60}$     & \{SFEDL~,~FEDC\}$\rightarrow$ \{2 CWNLs~,~ WNN\}\\
            sg.33             &  8n                                 & $\{a,b,c\}_{sg.33}=\{c,b,-a\}_{SG.60}$    & \{SFEDL~,~FEDC\}$\rightarrow$ CWNCz\\
       \hline\hline
  \end{tabular}
    \begin{tablenotes}
        \footnotesize
        \item[1] SFEDLs/FEDC means the FEDL in SG.60 is composed of two separated FEDLs or only one Dirac chain, respectively.
        \item[2] DNLx/DNLz denotes Dirac nodal loop in $k_{x}=\pi$/$k_{z}=\pi$ plane, which will be reserved under SG.60-sg.14-class a/b transition, respectively.
        \item[3] CWNCx and CWNCz denote concentric Weyl nodal chain in the $k_{x}=\pi$ and $k_{z}=\pi$ plane, respectively, while WNN denotes the Weyl nodal net schematized in Fig.~\ref{fig:sg60-sm}(c).
      \end{tablenotes}
\label{table:pt-SG60}
\end{table*}

\begin{figure*}[th]
\begin{center}
\includegraphics[width=0.95\textwidth]{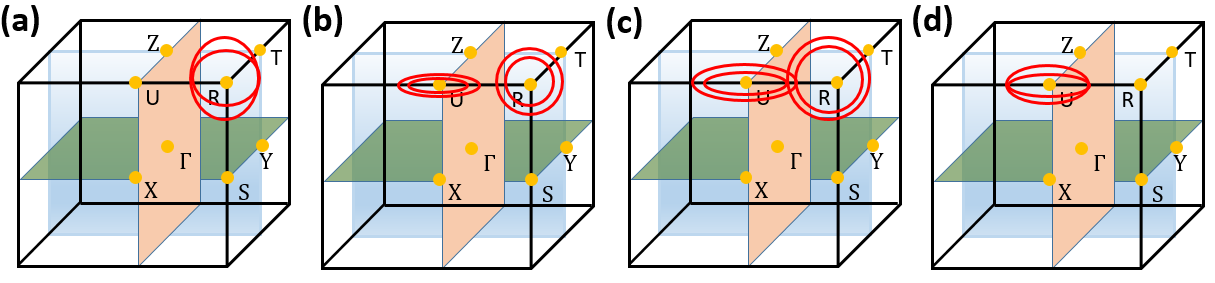}
\caption{%
  The evolution of FEDLs in SG.60 under (a) SG.60-sg.29, (b,c) SG.60-sg.30 and (d) SG.60-sg.33 transitions. The red loops denote how the FEDLs split under these transitions.
}
\label{fig:sg60-sm}
\end{center}
\end{figure*}

\begin{table*}
\caption{The symmetry-breaking transitions of SG.61.}
   \begin{tabular}{c c c}%
       \hline\hline
          \textbf{SG.61-sg.n} & \textbf{Filling factor of sg.n}     & \textbf{Phase transition}   \\
            sg.14-class $\beta$     &  4n                            & FEDL $\rightarrow$  DNL$\beta\oplus 2\mathcal{DP}\gamma\oplus 2\mathcal{DP}\alpha$\\
            sg.19                   &  4n                            & FEDL $\rightarrow$  $2\mathcal{DN}x\oplus2\mathcal{DN}y\oplus2\mathcal{DN}z$\\
            sg.29-class $\beta$     &  8n                            & FEDL $\rightarrow$  CWNC$\alpha$/WNLTBZ$\alpha$ $\oplus$ CWNL$\gamma$/WNCTBZ$\gamma$\\
       \hline\hline
  \end{tabular}
      \begin{tablenotes}
        \footnotesize
        \item[1] $2\mathcal{DP}\gamma$ and $2\mathcal{DP}\alpha$ mean that two DPs along the $k_{\alpha}=\pi \cap k_{\gamma}=\pi$ line are possible to exist when some FEDL protected by $\tilde{R}_{2\beta}$ symmetry crosses this line.
        \item[2] Similarly, $2\mathcal{DN}\alpha$ ($\alpha\in\{x,y,z\}$) denotes 2 DNs are possible to exist when 
            the FEDL confined in/traversing the BZ in the $k_{\alpha}=\pi$ plane is simultaneously protected by $\tilde{R}_{2\gamma}$/$\tilde{R}_{2\beta}$ symmetry, respectively.
        \item[3] In the case of SG.61-sg.29-class $\beta$ transition, CWNC$\alpha$ and WNLTBZ$\alpha$ are short for concentric Weyl nodal chain and Weyl nodal loop traversing the BZ in the $k_{\alpha}=\pi$ plane, while CWNL$\gamma$ and WNCTBZ$\gamma$ are short for concentric Weyl nodal loop and Weyl nodal chain traversing the BZ in the $k_{\gamma}=\pi$ plane, respectively. ``$A \oplus B$'' indicates nodal phenomena $A$ and $B$ are independent with each other, and can be available simultaneously, while ``$A/B$'' indicates nodal phenomena $A$ and $B$ are alternative under some transition. Fig.~\ref{fig:sg61-sm} shows how the FEDLs evolve under SG.61-sg.29-class $\beta$ transition.
      \end{tablenotes}
\label{table:pt-SG61}
\end{table*}
Similarly, we can deduce the evolutions of the FEDLs in the other four SGs. Especially, in terms of SG.61, class \{a,b,c\} of both SG.61-sg.14 and SG.61-sg.29 transitions corresponds to the axis chosen as the special direction of the symmetry-breaking perturbation, and the analyses are completely the same for the three classes due to the cyclic permutation relations shown in Table.~\ref{table:space-group}. Furthermore, SG.205 can be seen as a cubic case of SG.61, suggesting that the three FEDLs are related to each other by the $R_{3[111]}$ symmetry. The analyses of them are very similar, as shown in Table~\ref{table:pt-SG61} and Table~\ref{table:pt-SG205}, respectively.

\begin{figure*}[th]
\begin{center}
\includegraphics[width=0.95\textwidth]{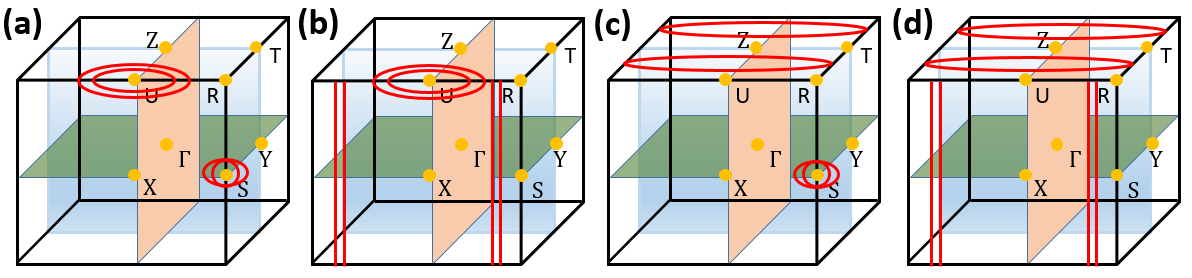}
\caption{%
  The evolution of the FEDLs in SG.61 under SG.61-sg.29-class b transition.
}
\label{fig:sg61-sm}
\end{center}
\end{figure*}

\begin{table*}
\caption{The symmetry-breaking transitions of SG.62.}
   \begin{tabular}{c c c c}%
       \hline\hline
          \textbf{SG.62-sg.n} & \textbf{Filling factor of sg.n}    & \textbf{Coordinate transformation}        & \textbf{Phase transition}   \\
            sg.11             &  4n                                 & $--$                                      & CFEDL$\rightarrow$ I\\
            sg.14-class a     &  4n                                 & $\{a,b,c\}_{sg.14}=\{-b,a,b+c\}_{SG.62}$  & FEDL$\rightarrow$ DNL\\
            sg.14-class b     &  4n                                 & $\{a,b,c\}_{sg.14}=\{b,c,a\}_{SG.62}$     & FEDL$\rightarrow$ 2$\mathcal{DP}$s\\
            sg.19             &  8n                                 & $--$                                      & FEDL$\rightarrow$ 2 $\mathcal{DN}$s\\
            sg.26             &  4n                                 & $\{a,b,c\}_{sg.26}=\{b,c,a\}_{SG.62}$     & FEDL$\rightarrow$ I\\
            sg.31             &  4n                                 & $\{a,b,c\}_{sg.31}=\{-b,a,c\}_{SG.62}$    & FEDL$\rightarrow$ CDNL\\
            sg.33             &  8n                                 & $\{a,b,c\}_{sg.33}=\{a,-c,b\}_{SG.62}$    & FEDL$\rightarrow$ CDNC\\
       \hline\hline
  \end{tabular}
        \begin{tablenotes}
        \footnotesize
        \item[1] If the two crossing points of FEDL along R-S are protected by $\tilde{R}_{2z}$, two DPs/DNs (signed as 2$\mathcal{DP}$s/2$\mathcal{DN}$s) appear under SG.62-sg.14-class b/SG.62-sg.19 transition, respectively.

      \end{tablenotes}
\label{table:pt-SG62}
\end{table*}

\begin{table*}
\caption{The symmetry-breaking transitions of SG.205.}
   \begin{tabular}{c c c}%
       \hline\hline
          \textbf{SG.205-sg.n} & \textbf{Filling factor of sg.n}    & \textbf{Phase transition}   \\
            sg.61                   &  8n                            & \{CFEDLs,FEDLTBZs,DNN\} $\rightarrow$  \{CFEDLs,FEDLTBZs,AFEDLs\}\\
            sg.148                  &  2n                            & \{CFEDLs,FEDLTBZs,DNN\} $\rightarrow$  I\\
            sg.198                  &  8n                            & \{CFEDLs,FEDLTBZs,DNN\} $\rightarrow$  \{6$\mathcal{DN}$s,6$\mathcal{DN}$s,I\}\\
       \hline\hline
  \end{tabular}
        \begin{tablenotes}
        \footnotesize
        \item[1] CFEDLs/FEDLTBZs indicate that the FEDLs are confined in/traversing the BZ, respectively, while DNN represents the Dirac nodal net schematized in Fig.~\ref{fig:sg61-205}(e). AFEDLs indicate the DNN can be tuned to arbitrary configuration of FEDLs in SG.61 under SG.205-sg.61 transition.
        \item[2] 6$\mathcal{DN}$s indicate that six four-fold DNs related with each other by $R_{3[111]}$ symmetry may appear at three $\tilde{R}_{2\alpha}$ invariant lines, S-R, T-R and U-R, respectively, when the FEDLs are protected by the $\tilde{R}_{2\alpha}$ symmetries. Otherwise, these FEDLs will be gapped. Even though the FEDLs may be gapped under the SG.205-sg.198 transition, fillings that realize a band insulator remain unchanged to be $8n$, which indicates the system is still half-filled and semimetallic.
      \end{tablenotes}
\label{table:pt-SG205}
\end{table*}

Thus, we can obtain the evolutions of the FEDLs in all the five SGs under all these transitions, the results are listed in Table~\ref{table:pt-SG57}, Table~\ref{table:pt-SG60}, Table~\ref{table:pt-SG61}, Table~\ref{table:pt-SG62} and Table~\ref{table:pt-SG205}, respectively. Furthermore, some schematic figures are given to visualise these evolutions, as shown in Fig.~\ref{fig:sg60-sm} and Fig.~\ref{fig:sg61-sm}, respectively.

\section{ Potential TIs and TCIs from the FEDL materials }
In the above, we have discussed the phase transitions of the FEDL materials under symmetry-breaking but translation-invariant perturbations. One may further ask if the FEDL materials will become TI or TCI when the FEDLs are gapped under the corresponding perturbations. In this section, we list almost all the FEDL materials and the corresponding symmetry indicators when they become insulators under these perturbations.

Fortunately, we find these insulators from the FEDL materials can be depicted with the $(Z_{2},Z_{2},Z_{2};Z_{4})$ indexes, just the same as centrosymmetric systems in SG.2. The values of these indexes can always be calculated for the $8n+4$ occupied bands and keep unchanged after these perturbations are applied. It is based on the adiabatic assumption that even though the degeneracies of the bands are violated under the perturbations, no new band crossings occur and the order of the bands keeps unchanged. The formula to calculate these indexes is expressed as 
\begin{equation}\label{eq:z2224}
\begin{split}
&Z_{2,1}\equiv \sum_{\mathbf{K} \in \mathbb{T R I M} \atop \text { at }\left\{k_{x}=\pi\right\}} \frac{N_{-}(\mathbf{K})-N_{+}(\mathbf{K})}{2}~mod~2,\\
&Z_{2,2}\equiv \sum_{\mathbf{K} \in \mathbb{T R I M} \atop \text { at }\left\{k_{y}=\pi\right\}} \frac{N_{-}(\mathbf{K})-N_{+}(\mathbf{K})}{2}~mod~2,\\
&Z_{2,3}\equiv \sum_{\mathbf{K} \in \mathbb{T R I M} \atop \text { at }\left\{k_{z}=\pi\right\}} \frac{N_{-}(\mathbf{K})-N_{+}(\mathbf{K})}{2}~mod~2,\\
&Z_{4}\equiv \sum_{\mathbf{K} \in \mathbb{T R I M} } \frac{N_{-}(\mathbf{K})-N_{+}(\mathbf{K})}{2}~mod~4.\\
\end{split}
\end{equation}
Using the above formula, we list the corresponding $(Z_{2},Z_{2},Z_{2};Z_{4})$ indexes of almost all the FEDL materials in each SG, as shown in Table~\ref{table:sg57-materials}, Table~\ref{table:sg60-materials}, Table~\ref{table:sg61-materials} and Table~\ref{table:sg205-materials}, respectively.
\begin{table*}
\begin{ruledtabular}
\caption{FEDL materials in SG.57.}
   \begin{tabular}{c c c c}%
          \textbf{Materials}    & \textbf{Materials}    & \textbf{Materials}      &        \textbf{Materials}         \\
            Cs(S;100,1)             &    Pr$_3$(GaNi)$_2$(S;000,2)   &   ReN$_2$(S;000,3)                &    YbCrSb$_3$(S;000,3)          \\
            K$_2$SnBi(S;000,0)         &    Pr$_3$(GeRu)$_2$(S;000,1)   &   K$_5$Hg$_7$(S;000,0)               &    ThTl(T;000,1)                \\
            CaAlPd(S;100,2)         &    La$_3$(GaNi)$_2$(S;100,1)   &   HfGa(S;000,1)                &    Na$_8$In$_6$Au$_{11}$(S;000,3)             \\
            Ca$_4$MgAl$_3$(S;000,1)       &     Y$_3$(SiRh)$_2$(S;000,1)   &    ---                         &      ---                 \\
  \end{tabular}
        \begin{tablenotes}
        \footnotesize
        \item[1] S/T is short for separated/traversing, which indicates the configuration of the FEDLs in SG.57 is composed by two separated FEDLs or two FEDLs traversing the BZ, respectively.
       \end{tablenotes}
\label{table:sg57-materials}
\end{ruledtabular}
\end{table*}

\begin{table*}
\begin{ruledtabular}
\caption{FEDL materials in SG.60.}
   \begin{tabular}{c c c c}%
          \textbf{Materials}    & \textbf{Materials}    & \textbf{Materials}    &        \textbf{Materials}         \\
            Fe$_2$N(S;C;110,0)     &       ReO$_2$(C;C;110,2)        &    RbCuCl$_3$(C;C;110,2)       &         Al$_3$FeSi$_2$(S;C;110,3)   \\
            Mn$_2$N(S;C;110,3)     &       Si$_2$Ni$_7$P$_5$(C;C;110,1)        &    Hf$_2$Co$_3$Si$_4$(C;C;110,3)          &         ---    \\
  \end{tabular}
  \begin{tablenotes}
        \footnotesize
        \item[1] S/C means that the FEDLs are composed by two separated Dirac loops or only one Dirac chain, while C/T is short for confined/traversing, which indicates the FEDL in the $k_{x}=\pi$ plane is confined in or traversing the BZ, respectively.
      \end{tablenotes}
\label{table:sg60-materials}
\end{ruledtabular}
\end{table*}

\begin{table}
\begin{ruledtabular}
\caption{FEDL materials in SG.61.}
   \begin{tabular}{c}%
          \textbf{Materials}       \\
            AgF$_2$(111;2)                \\
  \end{tabular}
    \begin{tablenotes}
        \footnotesize
        \item[1]  There exists only one FEDL material in SG.61 from the Materials Project. The configuration of the FEDLs contains an hourglass Dirac chain traversing the BZ. More details can be obtained from the earlier work~\cite{Shao-AgF2-2019}.
      \end{tablenotes}
\label{table:sg61-materials}
\end{ruledtabular}
\end{table}

\begin{table*}
\begin{ruledtabular}
\caption{FEDL materials in SG.205.}
   \begin{tabular}{c c c c}%
          \textbf{Materials} & \textbf{Materials}    & \textbf{Materials}            &        \textbf{Materials}         \\
            CuS$_2$(C;111,2)    &      Te$_2$Ir(C;111,3)            &    AuN$_2$(C;111,0)     &         IrN$_2$(N;111,0)   \\
            IrS$_2$(C;111,3)    &      CuTe$_2$(C;111,3)            &    CoTe$_2$(C;111,3)    &         CoS$_2$(C;111,2)    \\
            CoSe$_2$(C;111,2)   &      RhS$_2$(C;111,3)             &    CuSe$_2$(C;111,2)    &         Sb$_2$Au(C;111,3)    \\
            Te$_2$Rh(C;111,3)   &      RhSe$_2$(C;111,3)            &    --            &         --   \\
  \end{tabular}
      \begin{tablenotes}
        \footnotesize
        \item[1]  C/T/N is short for confined/traversing/net, which indicates that the FEDLs consist of three separated FEDLs confined in/traversing the BZ, or only one whole Dirac nodal net, respectively. The first-principles calculations indicate IrN$_{2}$ is the only material possessing the Dirac nodal net as schematized in Fig.~\ref{fig:sg61-205}(e).
      \end{tablenotes}
\label{table:sg205-materials}
\end{ruledtabular}
\end{table*}

For the case of FEDL materials in SG.62, we have listed the corresponding Z$_4$ index in the supplementary materials, with the three other Z$_2$ indexes $Z_{2x}\equiv Z_{2z}\equiv 1,Z_{2y}\equiv 0$. The values of these three Z$_2$ indexes are always fixed, which can be seen from the following. We find only bands at U contribute to the $Z_{2x}(=Z_{2z})$ index, because bands are always four-fold degenerate with the parities $p=2\times(+1)\oplus2\times(-1)$ at the other TRIMs in the $k_{x}=\pi$ and $k_{z}=\pi$ planes. 
Furthermore, bands are four-fold degenerate with the parities $p=4\times(+1)$ or $p=4\times(-1)$ at U, 
which indicates $N_{-}(\mathbf{U})-N_{+}(\mathbf{U})\equiv 2+4m$ for the fillings of $8n+4$, as a result, $Z_{2x}=Z_{2z}=1$. In addition, the four TRIMs in $k_{y}=\pi$ plane are all four degenerate with the parities $p=2\times(+1)\oplus2\times(-1)$, which gives $Z_{2y}=0$.

\section{ Two examples: $Pbca$ $AgF_2$ and $Pnma$ $SrIrO_3$ }
$Pbca$ AgF$_2$ is an interesting FEDL material, which has been studied in the earlier work~\cite{Shao-AgF2-2019}.
In the following, we will illustrate how this FEDL material evolves to various topological semimetals.

\begin{figure*}[th]
\begin{center}
\includegraphics[width=0.95\textwidth]{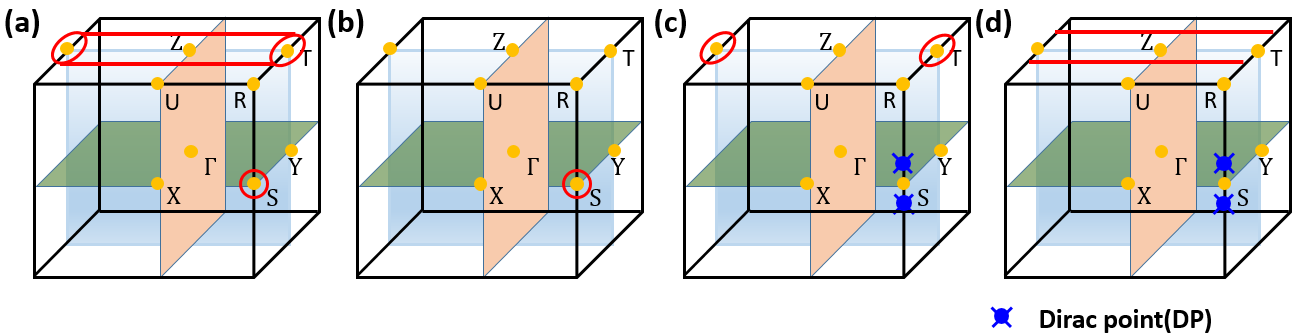}
\caption{%
  (a) The schematic figure of the FEDLs in $Pbca$ AgF$_2$.
  The evolution of FEDLs in $Pbca$ AgF$_2$ under (b) SG.61-sg.14-class a (c) SG.61-sg.14-class b and (d) SG.61-sg.14-class c transitions, respectively.
  The blue discs along R-S represent Dirac points originating from the splitting of the FEDL.
}
\label{fig:Pbca-AgF2-sg14}
\end{center}
\end{figure*}

\begin{figure*}[th]
\begin{center}
\includegraphics[width=0.95\textwidth]{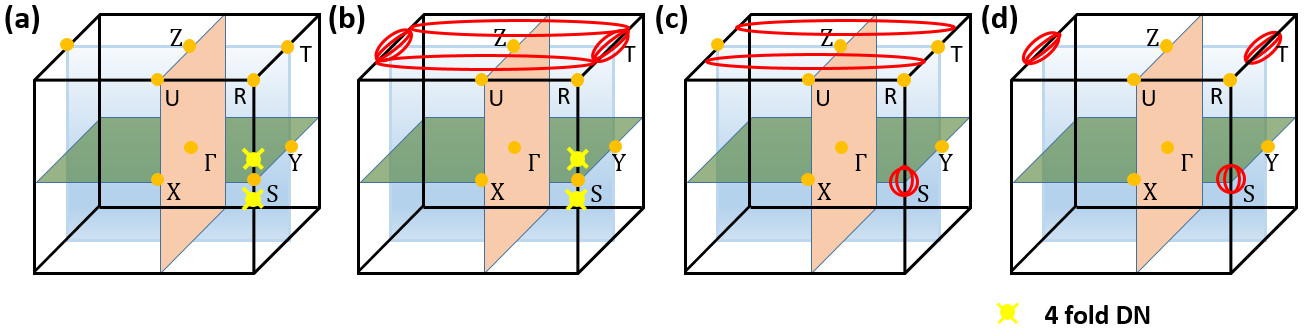}
\caption{%
  The evolution of FEDLs in $Pbca$ AgF$_2$ under (a) SG.61-sg.19, (b) SG.61-sg.29-class a, (c) SG.61-sg.29-class b and (d) SG.61-sg.14-class c transitions, respectively.
  The yellow discs originating from the splitting of the FEDL in the $k_{x}=\pi$ plane denote the four-fold degenerate nodes (DNs) along R--S.
}
\label{fig:Pbca-AgF2-sg19-29}
\end{center}
\end{figure*}

According to Eq.~(\ref{eq:SG61-zxyz}), we get $\{\zeta_{2x},\zeta_{2y},\zeta_{2z}\}=\{1,1,0\}$ from the first-principles calculations, which indicates the FEDLs in the $k_{x}=\pi~(k_{y}=\pi)$ and $k_{z}=\pi$ planes are confined in and traversing the BZ, respectively. Combined with the band structures shown in Fig.~\ref{fig:sg57-62-band} (c), we find the FEDL confined in the $k_{x}=\pi$ plane touches with the FEDL traversing the $k_{z}=\pi$ plane at some point along R-M$^{\prime\prime}$, leading to a Dirac chain traversing the BZ, as shown in Fig.~\ref{fig:Pbca-AgF2-sg14} (a).
Using Table.~\ref{table:pt-SG61}, and keep in mind that the node along the R-T line from the Dirac chain is not protected by $\tilde{R}_{2x}$ symmetry, while the node along R-S from the FEDL is protected by $\tilde{R}_{2z}$ symmetry, we can easily deduce the final nodal phenomena under all the SG.61-sg.n symmetry-breaking transitions, as shown in Fig.~\ref{fig:Pbca-AgF2-sg14} and Fig.~\ref{fig:Pbca-AgF2-sg19-29}, respectively.

\begin{figure}[bh]
\begin{center}
\includegraphics[width=0.45\textwidth]{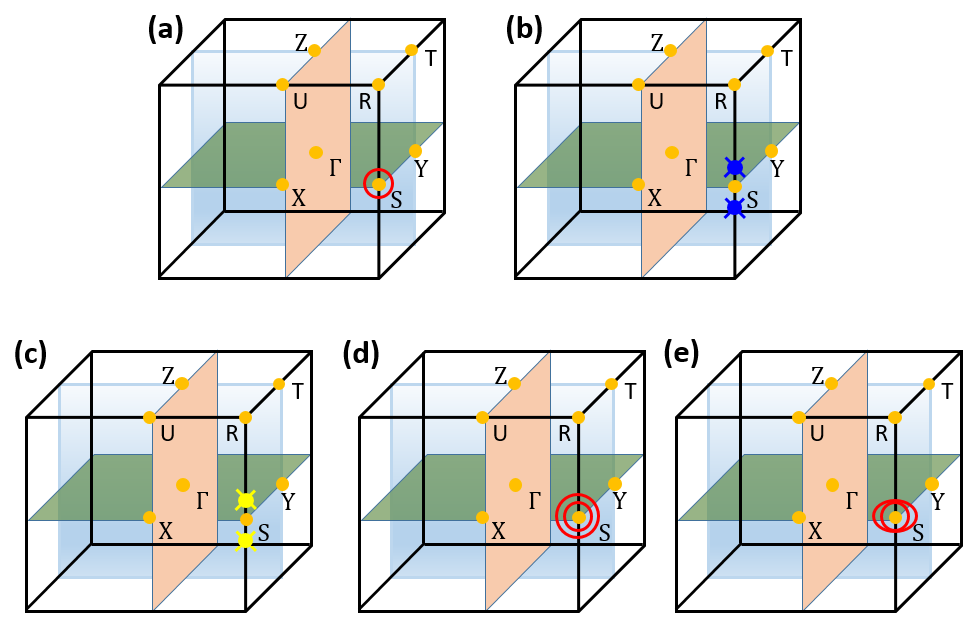}
\caption{%
  The evolution of FEDLs in $Pnma$ SrIrO$_3$ under (a) SG.62-sg.14-class a, (b) SG.62-sg.14-class b, (c) SG.62-sg.19, (d) SG.62-sg.31 and (e) SG.62-sg.33 transitions, respectively.}
\label{fig:SrIrO3}
\end{center}
\end{figure}

$Pnma$ SrIrO$_3$ is the well-known perovskite-class material which possesses the FEDL state~\cite{Kee-2012,Kee-2015}. We also deduce the evolutions of the FEDL under all maximal-subgroup transitions which respect the translation symmetries, as shown in Fig.~\ref{fig:SrIrO3}.

\section{ CONCLUSIONS }
We propose the FEDL state in non-magnetic systems and find it exists in five and only five SGs (SG.57, SG.60, SG.61, SG.62 and SG.205). Then we explore the possible configurations of the FEDLs in each SG. Band structures of $Pbcm$ K$_2$SnBi, $Pbcn$ Hf$_2$Co$_3$Si$_4$, $Pbca$ AgF$_2$, $Pnma$ AgAsF$_7$ and $Pa\bar{3}$ IrN$_2$ have been calculated to show the FEDLs in each SG.
Furthermore, we study the evolutions of the FEDLs under various perturbations which respect the translation symmetries, and find that the FEDL materials can serve as perfect parent materials of both topological semimetals with nodal points/loops, and TIs/TCIs.
At last, almost all the FEDL materials are listed, among which we have chosen $Pbca$ AgF$_2$ and the well-known perovskite-class material  $Pnma$ SrIrO$_3$ as two examples to illustrate how we can obtain various topological semimetals from the FEDL materials.

\section{ ACKNOWLEDGMENTS }
We thank for the fruitful discussions with Huaiqiang Wang, Jiachen Gao, Zhijun Wang, Zhaopeng Guo and Jiawei Ruan.

\bibliography{FEDL}

\end{document}